\documentclass[iop]{emulateapj}

\usepackage{graphicx}
\usepackage[active]{srcltx}
\usepackage{url}
\usepackage{multirow}
\usepackage{amssymb}
\usepackage{amsmath}
\usepackage{lineno}

\renewcommand{\d}{\mbox{d}}

\begin{document}

\title{Sensitivity of stacked imaging detectors to hard X-ray polarization}

\author{Fabio Muleri and Riccardo Campana}
\affil{INAF/IAPS, Via del Fosso del Cavaliere 100, I-00133 Roma, Italy}
\email{fabio.muleri@iaps.inaf.it}
\shorttitle{Sensitivity of stacked imaging detectors to hard X-ray polarization}
\shortauthors{Muleri \& Campana}

\begin{abstract}
The development of multi-layer optics which allow to focus photons up to 100~keV and more promises
an enormous jump in sensitivity in the hard X-ray energy band. This technology is already planned to
be exploited by future missions dedicated to spectroscopy and imaging at energies $>$10~keV, e.g.
\emph{Astro-H} and \emph{NuSTAR}. Nevertheless, our understanding of the hard X-ray sky would
greatly benefit from carrying out contemporaneous polarimetric measurements, because the study of
hard spectral tails and of polarized emission often are two complementary diagnostics of the same
non-thermal and acceleration processes. At energies above a few tens of keV, the preferred technique
to detect polarization involves the determination of photon directions after a Compton scattering.
Many authors have asserted that stacked detectors with imaging capabilities can be exploited for
this purpose. If it is possible to discriminate those events which initially interact in the first
detector by Compton scattering and are subsequently absorbed by the second layer, the direction of
scattering is singled out from the hit pixels in the two detectors. In this paper we give the first
detailed discussion of the sensitivity of such a generic design to the X-ray polarization. The
efficiency and the modulation factor are calculated analytically from the geometry of the
instruments and then compared with the performance as derived by means of Geant4 Monte Carlo
simulations.
\end{abstract}

\section{Introduction}

After the first pioneering experiments in the '70s \citep{Novick1972,Weisskopf1976,Weisskopf1978},
only very few polarimetric measurements have been carried out in high energy astronomy
\citep{Coburn2003,Rutledge2004,Willis2005,Kalemci2007,Dean2008,Forot2008,Yonetoku2011b}. The main
reason which prevented polarimetry to become a common tool also in this energy band was that even
state-of-art instruments were able to measure the polarization of only the brightest X-ray sources.
In the soft X-ray energy range, where grazing incidence optics were available, Bragg diffraction
polarimeters allowed only for a modest quantum efficiency, whereas Thomson scattering polarimeters
had a energy threshold mismatched with the energy band pass of the telescopes \citep{Novick1975}. At
higher energies, without the possibility to focus X-rays, Compton polarimeters required a large
collecting area and consequently the high background limited the sensitivity. As a result, no
dedicated polarimeters after the one on-board the OSO-8 satellite were launched, with the exception
of the small polarimeter \emph{Gamma-Ray Burst Polarimeter} (GAP) recently launched on-board the
Japanese satellite IKAROS to observe with a large field of view the prompt emission of Gamma Ray
Bursts \citep{Yonetoku2011}.

Today polarimetry is a field of growing interest in high energy astrophysics \citep{xraypol2010}.
At low energy, gas detectors able to image the path of the photoelectron in low atomic number
mixtures, e.g. Helium or Neon and dimethyl ether (DME), are a valuable alternative to Bragg
diffraction and Thomson scattering polarimeters \citep{Costa2001,Black2007,Bellazzini2010}. The
mission \emph{Gravity and Extreme Magnetism SMEX} (GEMS), a small explorer satellite planned to be
launched in 2014, exploits these photoelectric polarimeters together with grazing incidence
telescopes and promises to perform polarimetry of sources as faint as a few mCrab between 2 and
10~keV, with an enormous improvement in sensitivity with respect to OSO-8 polarimeters
\citep{Black2010,Jahoda2010}. On the other hand, the recent development of multi-layer optics
\citep{Christensen1992,Pareschi2003} makes attractive to extend this energy range upward. The
energy range of photoelectric polarimeters can be extended above 10~keV by using higher atomic
number mixtures, like Argon and DME possibly at high pressure \citep{Muleri2006, Soffitta2010}, but
above a few tens of keV Compton polarimeters \citep{McConnell2010} become more appealing because of
the prevailing probability of Compton scattering with respect to photoabsorption.

The design of a Compton polarimeter is of the greatest importance both to achieve the best
sensitivity and to reduce the systematic effects. Indeed, there are a number of very different
proposals in the literature \citep[see e.g.][]{Krawczynski2011}. In this paper we want to discuss
the performance of a particular design which, although not dedicated to polarimetry, can
provide some polarimetric sensitivity contemporaneously with imaging and spectroscopy. As a matter
of fact, many authors have asserted that such a design would endow next imaging/spectroscopy X-ray
missions with the capability to detect also polarization from bright sources
\citep{Barret2003,Gouiffes2008,Ferrando2010}. The geometry of such an instrument will be described
in Section~\ref{sec:Polarimeter}, while the performance will be investigated with a full analytical
treatment and with Geant4 simulations in Section~\ref{sec:Analytical} and in Section~\ref{sec:MC},
respectively.

\section{Description of the stacked imager polarimeter} \label{sec:Polarimeter}

\begin{figure*}[htbp]
\centering
\includegraphics[width=12cm]{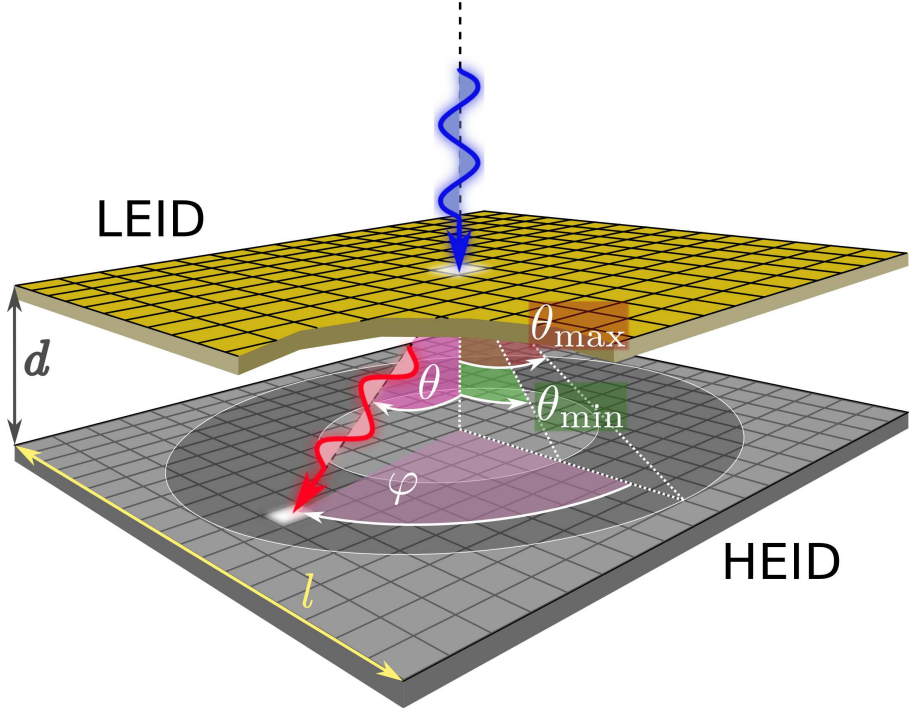}
\caption{Layout of the polarimeter studied in this paper. Polarization of photons is measured by
selecting those events which are scattered by the first detection plane (named LEID, Low Energy
Imaging Detector) and subsequently absorbed by the High Energy Imaging Detector (HEID). The
scattering angle $\theta$ and the azimuthal direction $\varphi$ are singled out by joining the
pixels hit in the LEID and in the HEID. The accepted events have to be scattered in the hollow cone
defined by $\theta_{\mathrm{min}}$ and $\theta_{\mathrm{max}}$, with
$\theta_{\mathrm{min}}\in[0,\theta_{\mathrm{max}})$ and
$\theta_{\mathrm{max}}<\arctan\frac{l}{2d}$.}
\label{fig:ComptonImager}
\end{figure*}

A common instrument layout dedicated to broad-band imaging in the focal plane of a X-ray
telescope exploits two stacked pixelated detection planes designed to absorb photons at different
energies (e.g. the imager camera on-board NHXM, \cite{Catalano2010}, or the Wide Field Imager
on-board the International X-ray Observatory, \cite{Stefanescu2010}). The first detector is 
usually made of Silicon and it is used to detect photons at lower energy ($\lesssim$1 to
$\sim$15~keV, depending on the thickness of the depletion layer), while harder photons, up to
$\sim$100~keV, are absorbed in the second detector, which is based on CdTe or CZT crystals.
Although the largest part of the high energy photons passes through the first detection plane
without interacting, a tiny fraction is absorbed or scattered by Silicon. For what concerns this
paper, only the latter events are those interesting for measuring polarization. If the energy
deposit due to the scattering in the Silicon is above the detection threshold and, after the
scattering, the photons are absorbed by the second detection plane, the direction of scattering can
be singled out by joining the hit pixels in the two detectors (see Figure~\ref{fig:ComptonImager}).

Therefore, any instrument with two stacked imaging detectors has, besides imaging capabilities, also
an intrinsic sensitivity to polarization. To exploit it, one must distinguish among the events
which release an energy deposit in both detectors, that we call ``double events'', those in which
the photon underwent a scattering in the first detection plane and was absorbed in the second one. A
first distinctive feature of such ``good'' events is that the signals from the two detection planes
are basically contemporaneous and therefore the LEID and the HEID must be put into coincidence. The
time window must be short enough to make negligible the probability of accidental coincidences due
to, for example, background or pile-up and this may require a coincidence window of a few $\mu$s or
less, and in any case a shaping time $\approx\mu$s. Another constraint for an event to be a good
Compton double event is that the energy released in the first detector must be below a certain
threshold corresponding to the maximum accepted scattering angle. The threshold is below a few keV
for incident photons with energy less than 40~keV and reasonable values of the scattering angle. 

The implementation of these constraints poses a number of requirements on the two detectors. Some
of them, e.g. the fast read-out, are quite severe for the Silicon-based detection plane. Nonetheless
they seem to be in the range of the current technology based on Silicon Drift Detectors
\citep{Lechner2010}. In the following we will not discuss in more detail the actual feasibility of
the instrument taken as an example in this paper, because our primary aim is to assess the
\emph{intrinsic} sensitivity to polarization of the stacked imager design. Therefore, we will
naively assume a ``toy model'' including only the fundamental components. The first pixelated
detection plane, hereafter called Low Energy Imaging Detector (LEID), is made of Silicon, while the
second stage is made of CdTe crystals and will be referred to as High Energy Imaging Detector
(HEID). The characteristics assumed for the LEID and for the HEID are reported in
Table~\ref{tab:LEID_HEID} and they are in line with those actually required in modern mission
proposals. We will also suppose that the time coincidence between the two detectors is fast enough
that the occurrence of spurious coincidences is negligible, as the background. Since the instrument
discussed is designed to be used in the focal plane of a telescope, we will focus our attention in
the energy range 20--100~keV, which is the interval where multilayer optics will be applied in the
next years.

\begin{table}[htbp]
\centering
\caption{Characteristics of the stacked imager polarimeter assumed throughout this paper.}
\label{tab:LEID_HEID}
\begin{tabular}{ll}
\tableline\noalign{\smallskip}
\multicolumn{2}{l}{\bf Low Energy Imaging Detector (LEID)}\\
\tableline\noalign{\smallskip}
Material & Silicon \\
Thickness & 450~$\mu$m \\
Area & 51.2$\times$51.2~mm$^2$\\
Pixels & 512$\times$512 pixels, square pattern\\
Pixel size & 100~$\mu$m$\times$100~$\mu$m\\
\tableline\noalign{\smallskip}
\multicolumn{2}{l}{\bf High Energy Imaging Detector (HEID)}\\
\tableline\noalign{\smallskip}
Material & CdTe \\
Thickness & 2~mm \\
Area & 51.2$\times$51.2~mm$^2$\\
Pixels & 256$\times$256 pixels, square pattern\\
Pixel size & 200~$\mu$m$\times$200~$\mu$m\\
\tableline\noalign{\smallskip}
Distance LEID/HEID & 2~cm\\
Energy range & 20--100~keV \\
\tableline\noalign{\smallskip}
\end{tabular}
\end{table}

In the following, it will be implicitly assumed that the Minimum Detectable Polarization, that is
the upper limit to the polarization which can be statistically measured in a certain observation
time and with a certain confidence, is fully representative of the sensitivity of the instrument.
This approach is purely statistical and, consequently, it will neglect a very important requirement
of any actual polarimeter, i.e. the reduction to a negligible level and, if this is not possible,
the correction of any systematic effects which may result in a spurious polarized signal.
Systematics may be severe for the polarimeter discussed here, because the probability of scattering
in the first detection plane is tiny and therefore there are a number of small effects which may
play a role. Without the intention to be complete, systematic effects may arise from nonuniformities
in the pixel efficiency, energy threshold or geometry, or they may come out from pointing
instabilities or anisotropic background. These issues must be extensively studied if dealing with
the feasibility of the polarimeter design discussed in this paper, but, again, this is out of the
scope of this work where we rather focus on the ``bulk'' of the problem, that is the performance of
the method in ideal conditions.

\section{Sensitivity to polarization - Analytical treatment} \label{sec:Analytical}

Compton polarimeters detect the polarization of the incident radiation by measuring the photon
scattering direction: a polarized signal causes a square-cosine modulation to appear in the
histogram of azimuthal directions of scattering (hereafter \emph{modulation curve}). Even if the
incident radiation is completely unpolarized, the number of photons scattered per angular bin is
Poisson-distributed around the mean value, and this always mimics a spurious modulation to some
extent. The amount of this modulation at a certain confidence level translates in a Minimum
Detectable Polarization (MDP) and, by definition, only a detection greater than MDP is statistically
significant \citep{Weisskopf2010}.

The MDP is calculated from the source and the background fluxes in the selected energy range, $F$
and $B$ respectively, and for the 99\% confidence level it is \citep{Weisskopf2010}:
\begin{equation}
\mbox{MDP} =\frac{4.29}{\epsilon \mu F} \sqrt{\frac{B + \epsilon F}{S T}}\, ,
\label{eq:MDP}
\end{equation}
where $T$ is the observation time, $S$ the collecting area, $\epsilon$ is the detector efficiency
and $\mu$ the \emph{modulation factor}. This latter parameter is defined as the amplitude of the
modulation curve when completely polarized and monochromatic photons are incident on the instrument:
\begin{equation}
\mu = \frac{M_{\max}-M_{\min}}{{M}_{\max}+M_{\min}}\, ,
\label{eq:MuDef}
\end{equation}
where $M_{\max}$ and $M_{\min}$ are the maximum and the minimum of the modulation curve,
respectively. A higher value of $\mu$ means that the instrument responds to polarized radiation
with a larger modulation and the effect of statistical fluctuations is, in proportion, lower.

In our toy model, we assumed that spurious double events and background are negligible and,
therefore, the MDP is inversely proportional to the \emph{quality factor} $q=\mu\sqrt{\epsilon}$:
\begin{equation}
\mbox{MDP} \simeq\frac{4.29}{\mu \sqrt{\epsilon}} \sqrt{\frac{1}{F S T}}\propto\frac{1}{q}\, .
\label{eq:MDP_QF}
\end{equation}
In the following we will use the quality factor to optimize the sensitivity of the polarimeter: the
larger $q$ the higher the sensitivity. To evaluate $q$, we need to calculate $\mu$ and $\epsilon$
separately.

\subsection{Estimate of the modulation factor}

An estimate of the modulation factor can be computed from some very basic considerations, taking
into account the geometry of the instrument and the angular dependence of the differential cross
section of Compton scattering. The latter, for completely polarized photons incident on a free
electron at rest, is \citep{Klein1929,Heitler1954}:
\begin{equation}
\frac{\d\sigma_{KN}}{\d\Omega} = \frac{1}{2} r_0^2 \frac{E'^2}{E^2} \left[ \frac{E}{E'} +
\frac{E'}{E} -2\sin^2\theta\cos^2\phi \right] \, ,
\label{eq:KN}
\end{equation}
where $\theta$ is the scattering angle, $\phi$ is defined as the azimuthal angle from the
scattering direction to the polarization vector of the incident photon and $r_0=
\frac{1}{4\pi\epsilon_0}\frac{e^2}{m c^2}$ is the classical electron radius. The ratio between
the energy before and after the scattering ${E}/{E'}$ is related to the scattering angle $\theta$
by the Compton formula:
\begin{equation}
\frac{E}{E'} = 1+\varepsilon\left(1-\cos\theta\right) \, ,
\label{eq:EnCmp}
\end{equation}
where $\varepsilon$ is the energy of the incident photon in unit of electron mass,
$\varepsilon=E/m_e
c^2$. Equation~\ref{eq:KN}, by substituting Equation~\ref{eq:EnCmp}, becomes:
\begin{eqnarray}
\frac{\d\sigma_{KN}}{\d\Omega} &=& \frac{1}{2} r_0^2
\Bigg\{\frac{1}{1+\varepsilon\left(1-\cos\theta\right)} + \nonumber\\
&+&\frac{1}{
\left[1+\varepsilon\left(1-\cos\theta\right)\right]^3}
-\frac{2\sin^2\theta\cos^2\phi}{\left[1+\varepsilon\left(1-\cos\theta\right)\right]^2} \Bigg\} \, .
\end{eqnarray}
In the following we will be interested only in the angular dependence of the differential cross
section. Therefore we define the function $D(\varepsilon,\theta,\phi)$ as:
\begin{eqnarray}
D(\varepsilon,\theta,\phi) &=& \Bigg\{\frac{1}{1+\varepsilon\left(1-\cos\theta\right)} +\nonumber\\
&+&\frac{1}{\left[1+\varepsilon\left(1-\cos\theta\right)\right]^3} -\nonumber\\
&-&\frac{2\sin^2\theta\cos^2\phi}{\left[1+\varepsilon\left(1-\cos\theta\right)\right]^2} \Bigg\} \, .
\end{eqnarray}

The first step to evaluate the modulation factor is to calculate the modulation curve, i.e.
to make an histogram of the azimuthal scattering directions for completely polarized and
monochromatic photons. We have to count how many photons are scattered in a certain azimuthal
direction $\phi$, regardless of the scattering angle $\theta$. The number of photons $\d N
(\varepsilon,\theta,\phi)$ scattered in the $\phi$ and $\theta$ direction per unit of solid angle
$\d\Omega$ at energy $\varepsilon$ is proportional to the differential cross section:
\begin{equation}
\d N (\varepsilon,\theta,\phi) = \kappa \; D(\varepsilon,\theta,\phi) \sin\theta \d\theta \d\phi\, ,
\label{eq:dN}
\end{equation}
where $\kappa$ is a constant of proportionality and $\d\Omega=\sin\theta \d\theta \d\phi$.
Neglecting the probability of multiple scatterings in the LEID, the direction measured by joining
the hit pixels in the LEID and in the HEID will be that of scattering, and therefore the modulation
curve is obtained by summing ${\d N}/{\d\phi}$ over the range of $\theta$ values:
\begin{eqnarray}
M(\varepsilon,\varphi,\theta_{\mathrm{min}},\theta_{\mathrm{max}}) &=& \int_{\theta_{\mathrm{min}}}^{\theta_{\mathrm{max}}} \frac{\d
N(\varepsilon,\theta,\phi)}{\d\phi} =\nonumber\\
&=& \kappa \int_{\theta_{\mathrm{min}}}^{\theta_{\mathrm{max}}} 
D(\varepsilon,\theta,\phi) \sin\theta \d\theta \, .
\label{eq:ModCurve}
\end{eqnarray}
The limits of integration, i.e. $\theta_{\mathrm{min}}$ and $\theta_{\mathrm{max}}$, are at large
fixed by the geometry of the system. Referring to Figure~\ref{fig:ComptonImager}, and using the
parameters in Table~\ref{tab:LEID_HEID}, we have that in our case:
\begin{equation}
\theta_{\mathrm{max}} = \arctan\frac{l}{2d}\approx52^\circ\, ,
\label{eq:thetamax}
\end{equation}
where $l$ is the side of the HEID and $d$ is the LEID-HEID distance, and
$\theta_{\mathrm{min}}\in[0,\theta_{\mathrm{max}})$. We will discuss in
Section~\ref{sec:AnalyticalQF} how more stringent limits on $\theta$ affect the sensitivity to
polarization.

In the left hand side of Equation~\ref{eq:ModCurve} we used the variable $\varphi$ instead of $\phi$
because the latter is the angle to the photon polarization vector, the former is the angle to some
axis of reference of the instrument. The two angles are related by
\begin{equation}
\phi = \varphi-\varphi_0+\frac{\pi}{2}
\end{equation}
where $\varphi_0$ is the angle of polarization defined as the peak of the modulation curve. In the
following we will assume $\varphi_0=0$ and therefore the maximum of the modulation curve will be at
$\varphi=0$ (or $\phi=\pi/2$) and the minimum will be at $\varphi=\pi/2$ (or $\phi=0$). The
modulation factor is eventually derived by applying Equation~\ref{eq:MuDef}:
\begin{eqnarray}
&& \mu(\varepsilon,\theta_{\mathrm{min}},\theta_{\mathrm{max}}) = 
\frac{M_{\max}-M_{\min}}{{M}_{\max}+M_{\min}}=\nonumber\\
&=& \frac{M(\varepsilon,\varphi=0,\theta_{\mathrm{min}},\theta_{\mathrm{max}})-M(\varepsilon,
\varphi=\pi/2, \theta_{\mathrm{min}},\theta_\mathrm{max})}{M(\varepsilon,
\varphi=0,\theta_{\mathrm{min}},\theta_{\mathrm{max}})+M(\varepsilon,
\varphi=\pi/2,\theta_{\mathrm{min}},\theta_{\mathrm{max}})}=\nonumber\\
&=&  \int_{\theta_{\mathrm{min}}}^{\theta_{\mathrm{max}}}
\frac{D(\varepsilon,\theta,\pi/2)-D(\varepsilon,\theta,0)}{D(\varepsilon,\theta,\pi/2)+D(\varepsilon
, \theta, 0)} \sin\theta \d\theta \, .
\label{eq:Mu}
\end{eqnarray}

\begin{figure}[htbp]
\begin{center}
\includegraphics[width=8cm]{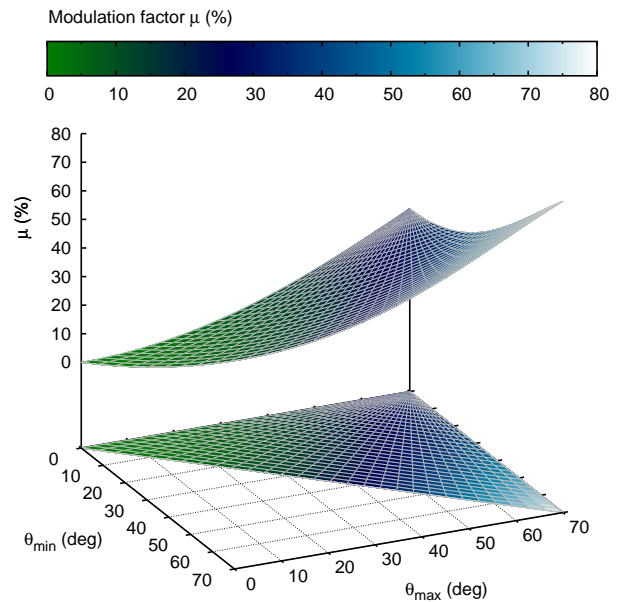}
\caption{Behavior of the modulation factor as a function of $\theta_{\mathrm{min}}$ and
$\theta_{\mathrm{max}}$. The energy of the incident photons is 30~keV, but, within the energy range
of our interest (20--100~keV), this does not affect the result significantly. We assumed that
$0\leq\theta_{\mathrm{min}}<\theta_{\mathrm{max}}<70^\circ$ to stress that our results are
not applicable for $\theta_{\mathrm{min}},\theta_{\mathrm{max}}\rightarrow90^\circ$.}
\label{fig:Mu_30keV}
\end{center}
\end{figure}

Equation~\ref{eq:Mu} can be integrated employing standard techniques, but the resulting explicit
expression is rather cumbersome and not easy to handle. Consequently, we study the behavior
of $\mu(\varepsilon,\theta_{\mathrm{min}},\theta_{\mathrm{max}})$ numerically, by means of the
Computer Algebra System \textsc{Maxima}\footnote{\url{http://maxima.sourceforge.net/}}. In
Figure~\ref{fig:Mu_30keV} we report the modulation factor as a function of $\theta_{\mathrm{min}}$
and $\theta_{\mathrm{max}}$. As an example, we set the energy of incident photons at 30~keV but the
qualitative behavior is fully representative, since the dependence of $\mu$ on the energy is rather
weak in the range of our interest. The value of the modulation factor increases monotonically with
$\theta_{\mathrm{max}}$ and $\theta_{\mathrm{min}}$ and, in principle, the maximum would reach
nearly 100\% for $\theta_{\mathrm{min}},\theta_{\mathrm{max}}\rightarrow90^\circ$. However, this
result is not relevant here because, if the scattering angle is close to 90$^\circ$, the assumption
of a single scattering in the LEID (which led to Equation~\ref{eq:ModCurve}) fails and so does our
model. Moreover, any feasible layout of the polarimeter discussed in this paper would require
$\theta_{\mathrm{min}}$ and $\theta_{\mathrm{max}}$ to be constrained well below 90$^\circ$. We
limited the range of $\theta_{\mathrm{max}}$ and $\theta_{\mathrm{min}}$ between 0 and 70$^\circ$ in
Figure~\ref{fig:Mu_30keV} to stress that our results are not applicable for
$\theta_{\mathrm{min}},\theta_{\mathrm{max}}\rightarrow90^\circ$.

\subsection{Estimate of the efficiency}
The efficiency $\epsilon$ of detecting scattered photons has to take into account three main
contributions: ($i$) the probability of Compton scattering in the LEID, ($ii$) the photoabsorption
efficiency in the HEID and ($iii$) that only a fraction of photons are actually scattered towards
the HEID. We evaluate the first term, $\epsilon_{\mathrm{scatt}}^{L}(E)$, as the fraction of Compton
scatterings among all of the interactions which occur in the LEID:
\begin{equation}
\epsilon_{\mathrm{scatt}}^L(E,x_L) = \frac{\mu_{\mathrm{scatt}}^L(E)}{\mu_{\mathrm{tot}}^L(E)}\;
\left[1-\exp{\left(-\mu_{\mathrm{tot}}^L(E)\;\rho^L\;x^L\right)}\right]\, ,
\label{eq:epsilon_scatt}
\end{equation}
where $\mu_{\mathrm{tot}}^L$ and $\mu_{\mathrm{scatt}}^L$ are the total and the incoherent
scattering attenuation coefficients for the LEID, $\rho^L$ is the density of the LEID material
(Silicon, in our model) and $x^L$ its thickness.

The second contribution to the efficiency takes into account the probability that the scattered
photon is absorbed in the HEID. We will assume in the following that the photoabsorption efficiency
of the HEID is 100\% in the energy range of our interest, i.e.:
\begin{equation}
\epsilon_{\mathrm{abs}}^H = 1\, .
\label{eq:epsilon_abs}
\end{equation}
This assumption, although naive, will not affect our results significantly. In effect, it is quite
reasonable because the requirement for the HEID to have an high efficiency in the entire energy
range is a heritage of its use as an imager. In the configuration assumed (see the
Table~\ref{tab:LEID_HEID}) the $x^{H} = 2$~mm thickness allows for the photoelectric absorption of
more that 80\% of 100~keV photons incident on-axis. Also, the efficiency in the polarimetric mode is
actually higher, because the scattered photons see the detector as having an effective depth
$x^H/\cos\theta$. Claiming a 100\% HEID efficiency allows both to simplify the following discussion
and to derive the maximum possible sensitivity to polarization.

The last contribution to the efficiency is due to the fact that, while photons are scattered over
all the solid angle, only in a few cases their scattering direction crosses the HEID. Actually, we
set a more stringent limit, i.e. we take only those photons scattered at angles between
$\theta_{\mathrm{min}}$ and $\theta_{\mathrm{max}}$, no matter the value of the azimuthal angle
$\phi$. The fraction of accepted photons $\epsilon_c$ is calculated integrating
$\d N(\varepsilon,\theta,\phi)$ (see Equation~\ref{eq:dN}):
\begin{eqnarray}
&&\epsilon_\mathrm{c}(\varepsilon,\theta_{\mathrm{min}},\theta_{\mathrm{max}}) 
= \frac{\displaystyle \int_0^{2\pi}\int_{\theta_{\mathrm{min}}}^{\theta_{\mathrm{max}}} \d N
(\varepsilon,\theta,\phi)}{\displaystyle \int_0^{2\pi} \int_{0}^{\pi} \d N
(\varepsilon,\theta,\phi)} =\nonumber\\
&=&  \frac{\displaystyle \int_0^{2\pi}\int_{\theta_{\mathrm{min}}}^{\theta_{\mathrm{max}}}
D(\varepsilon,\theta,\phi)
\sin\theta \d \theta \;\, \d\phi}{\displaystyle \int_0^{2\pi} \int_{0}^{\pi} 
D(\varepsilon,\theta,\phi) \sin\theta \d \theta \;\, \d\phi} \, .
\end{eqnarray}	

The efficiency of the polarimeter can be therefore evaluated from the three aforementioned
contributions:
\begin{eqnarray}
&&\epsilon(\varepsilon,\theta_{\mathrm{min}}, \theta_{\mathrm{max}},x^{L}) =\nonumber\\
&=&\epsilon_{\mathrm{scatt}}^L(\varepsilon,x_L)\times
\epsilon_\mathrm{abs}^H\times\epsilon_\mathrm{c}(\varepsilon,\theta_{\mathrm{min}},\theta_{\mathrm{
max}}) =
\nonumber\\
&=&\frac{\mu_{\mathrm{scatt}}^L(\varepsilon)}{\mu_{\mathrm{tot}}^L(\varepsilon)}\;
\left[1-\exp{\left(-\mu_{\mathrm{tot}}^L(\varepsilon)\;\rho^L\;x^L\right)}\right]\times \nonumber\\
&\times&\frac{\displaystyle
\int_0^{2\pi}\int_{\theta_{\mathrm{min}}}^{\theta_{\mathrm{max}}}
D(\varepsilon,\theta,\phi) \sin\theta \d \theta \;\, \d\phi}{\displaystyle \int_0^{2\pi}
\int_{0}^{\pi} D(\varepsilon,\theta,\phi) \sin\theta \d \theta \;\, \d\phi} \, ,
\label{eq:Eff}
\end{eqnarray}
where we used $\varepsilon=E/m_e c^2$ to express the dependence on the energy of the incident
photon.

\begin{figure}[htbp]
\begin{center}
\includegraphics[width=8cm]{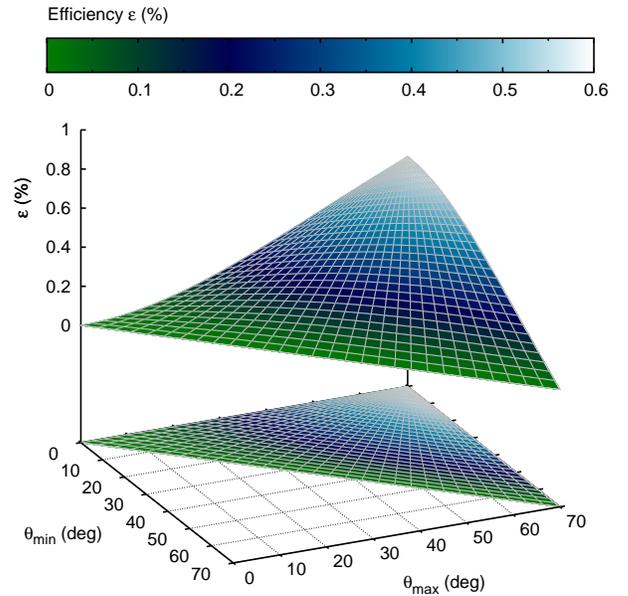}
\caption{Behavior of the efficiency as a function of $\theta_{\mathrm{min}}$ and
$\theta_{\mathrm{max}}$ for a LEID thickness of 450~$\mu$m. The energy of the incident photons is 
30~keV, but, within the energy range of our interest (20-100~keV), this does not affect the result
significantly.}
\label{fig:Eff_30keV}
\end{center}
\end{figure}

As for the modulation factor, we likewise plot in Figure~\ref{fig:Eff_30keV} the behavior of
$\epsilon(\varepsilon,\theta_{\mathrm{min}},\theta_{\mathrm{max}})$ as a function of
$\theta_{\mathrm{min}}$ and $\theta_{\mathrm{max}}$ for 30~keV photons. The qualitative result is
rather simple: the efficiency is higher when the interval of accepted scattering angles is
larger ($\theta_{\mathrm{max}}\rightarrow90^\circ$ and $\theta_{\mathrm{min}}\rightarrow0$). 
More remarkably, the absolute value of the efficiency  is always lower that 1\%. As a matter of
fact, the value of $\epsilon_{\mathrm{scatt}}^L$ is 1.6\% for the assumed thickness of the LEID
(450~$\mu$m) at 30~keV and the incomplete collection of the scattered photons makes things worse.
The low intrinsic efficiency is the most important deficiency in the design of a polarimeter based
on stacked imaging detectors and it cannot be easily overcome. The only way is to increase the
thickness of the depleted region of the LEID, but the value of $\epsilon_{\mathrm{scatt}}^L$ is only
3.4\% still in the case of a depleted region 1~mm thick, which is one of the largest values still
technologically reasonable. Moreover the increase of the thickness would result in a larger
background and would spoil the performance of the LEID for the detection of faint sources.
 
\subsection{Quality factor} \label{sec:AnalyticalQF}

Equations~\ref{eq:Mu} and \ref{eq:Eff} can be combined to obtain the quality factor
$q=\mu\sqrt{\epsilon}$. Notably, we enclosed the dependence of $q$ on the geometry of the detector
in only three parameters, that are $\theta_{\mathrm{min}}$, $\theta_{\mathrm{max}}$ and $x^{L}$.
This allows us to optimize the sensitivity of the polarimeter varying a very limited number of free
quantities. 

\begin{figure*}[htbp]
\begin{center}
{\includegraphics[width=10cm]{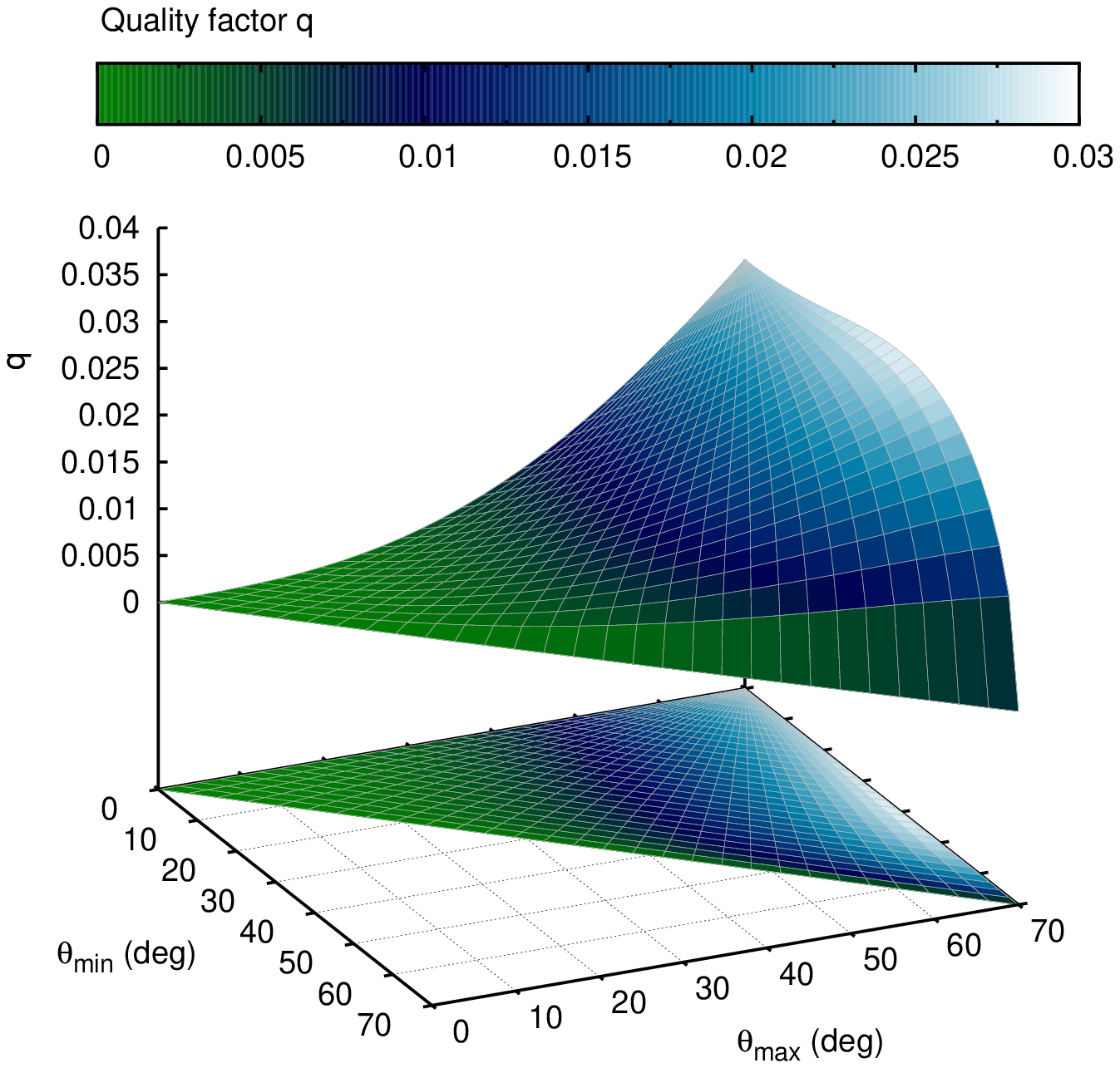}}
\hspace{5mm}
{\includegraphics[width=10cm]{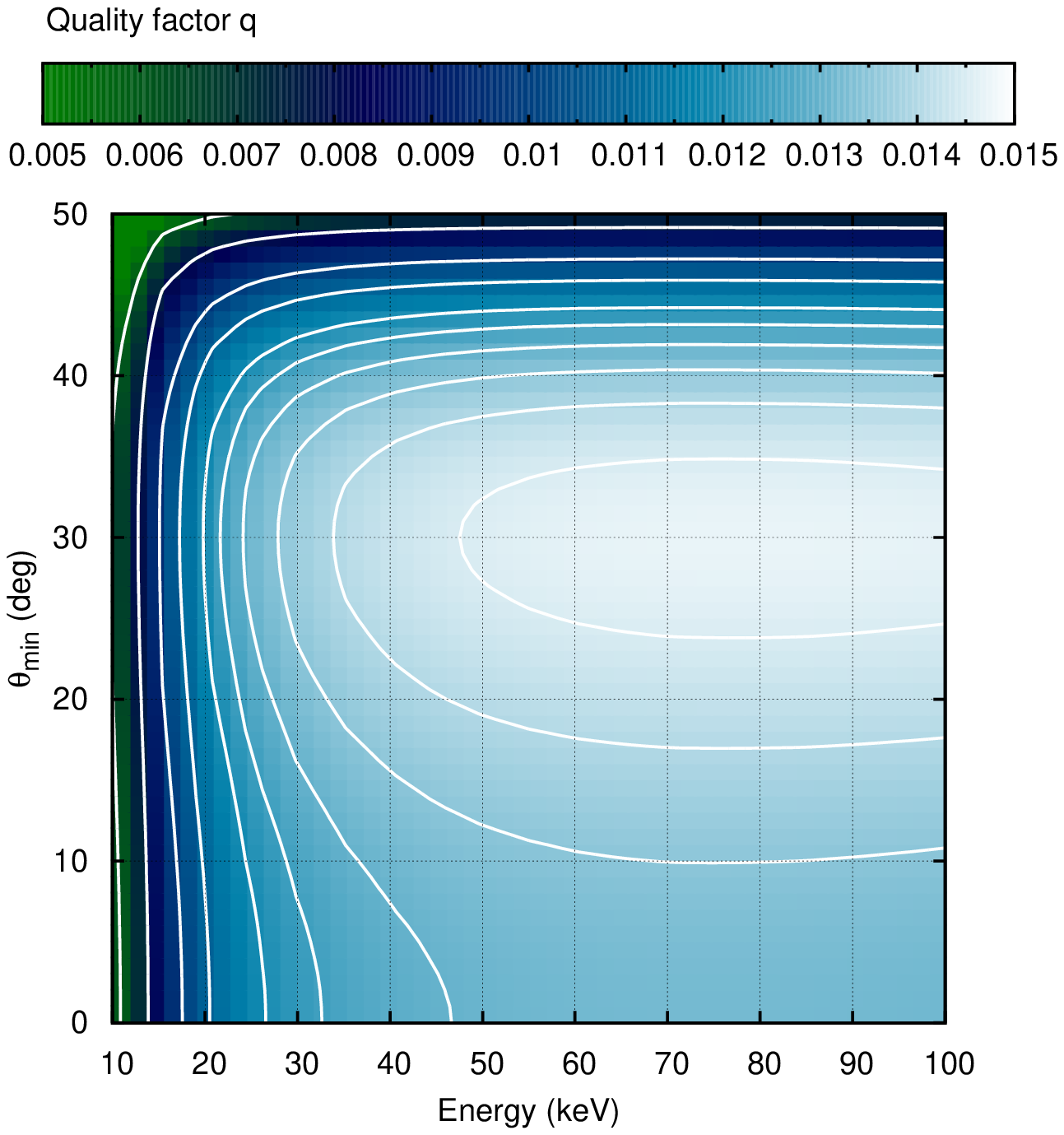}}
\end{center}
\caption{({\bf a}) Behavior of the quality factor as a function of $\theta_\mathrm{min}$ and
$\theta_\mathrm{max}$ at 30~keV. ({\bf b}) Quality factor as a function of $\theta_\mathrm{min}$ and
energy for $\theta_{\mathrm{max}} = 52^\circ$.}
\label{fig:QF_3}
\end{figure*}

The behavior of the quality factor at 30~keV with respect to $\theta_{\mathrm{min}}$ and
$\theta_{\mathrm{max}}$ is reported in Figure~\ref{fig:QF_3} for $x^{L}=450~\mu$m. As expected,
the sensitivity monotonically increases with $\theta_{\mathrm{max}}$ because both modulation factor
and efficiency do. In the geometry assumed in Table~\ref{tab:LEID_HEID},
$\theta_{\mathrm{max}}\lesssim52^\circ$ (cf. Equation~\ref{eq:thetamax}) and therefore we freeze
$\theta_{\mathrm{max}}$ to 52$^\circ$ in order to maximize $q$. On the contrary, the dependence on
$\theta_{\mathrm{min}}$ is more complex. The interplay between the increasing modulation factor and
the decreasing efficiency originate a broad peak in the quality factor. This unexpected result is
shown more clearly in Figure~\ref{fig:QF_3} where the dependence of $q$ with respect to
$\theta_{\mathrm{min}}$ and energy is reported for $\theta_{\mathrm{max}}=52^\circ$. The peak is
basically independent on energy in the range on our interest, but depends on 
$\theta_{\mathrm{max}}$. A practical relation which linkes $\theta_{\mathrm{max}}$ and the value of
$\theta_{\mathrm{min}}$ which maximizes $q$ is:
\begin{equation}
\theta_{\mathrm{min}}^{\mathrm{peak}} \mbox{\, [deg]} \approx 0.57\;\theta_{\mathrm{max}}\mbox{\,
[deg]}\, .
\label{eq:QFmax}
\end{equation}
In our case, $\theta_{\mathrm{max}}=52^\circ$ and therefore
$\theta_{\mathrm{min}}^{\mathrm{peak}}\approx30^\circ$. We will use these values in the following as
those which provide the best sensitivity to polarization, but the choice to accept only events
scattered with an angle larger than a certain threshold has also practical advantages. Events
scattered at angles lower than $30^\circ$ release in the LEID less than 0.23~keV at 30~keV according
to the Compton formula (see Equation~\ref{eq:EnCmp}), and such a low energy deposit may be difficult
to detect.

\section{Monte Carlo simulations}\label{sec:MC}

The results discussed in the previous section are purely analytical. Moreover, they rely on some
simplified physical assumptions, e.g. in treating the Compton scattering as from free electrons at
rest. In the general case of the scattering inside a realistic material, however, the momentum
distribution of the electrons and the binding effects in atoms will produce both a broadening of the
Compton peak in the scattered photon spectrum and a suppression of the scattering at forward angles
\citep{Veigele1966,Bergstrom1997,Hubbell1997}. The incoherent scattering cross section thus
becomes:
\begin{equation}
\frac{\d\sigma_\mathrm{inc}}{\d\Omega} = S_F(x, Z) \frac{\d\sigma_\mathrm{KN}}{\d\Omega}\, ,
\end{equation}
where $S_F(x, Z)$ is the so-called \emph{scattering function} and depends on the transferred
momentum $x\approx(E/hc)\sin(\theta/2)$ and on the atomic number $Z$ of the scattering atom. The
efficiency for a given scattering polar angle $\theta$ is therefore affected. The azimuthal angle
$\phi$, on the contrary, can be assumed independent of the binding effects \citep{Matt1996}. Another
effect to take into account is due to the ``pixelization'' of the detectors, and to the geometrical
shape of the pixel itself. To make a meaningful comparison with a realistic case, we therefore
developed a Monte Carlo simulator of our toy model for a stacked polarimeter (see
Table~\ref{tab:LEID_HEID}), using the Geant4 toolkit \citep{Agostinelli2003}, version 4.9.4.
Accurate physics lists were employed for the various electromagnetic interactions, notably the
Livermore low energy
libraries\footnote{\url{https://twiki.cern.ch/twiki/bin/view/Geant4/\\LoweMigratedLivermore}} that use
the EPDL97 tabulated version of the scattering function \citep{Hubbell1975}.

Geant4 simulations allow us to treat the physics of the scattering more accurately. Notwithstanding,
we intentionally maintain also in Monte Carlo treatment some simplified assumptions in the design of
the instrument. In particular, we do not apply any energy threshold to the event detection in the
LEID and consider LEID and HEID as having an infinite energy resolution. We are aware that these
parameters play an important role in the determination of a good estimate of the detection
efficiency, especially at low energy. In particular, any realistic assumption on the energy
threshold and on the spectral resolution would decrease the efficiency and hence the sensitivity to
polarization. Nonetheless, this choice is in line with our primary goal, that is to argue on the
intrinsic sensitivity to polarization of the stacked imager layout rather than to propose a feasible
design and to discuss its sensitivity.

\subsection{Data Analysis} \label{sec:DataAnlysis}

For each event (i.e. for each primary photon generated in the simulation run), the output of the
Geant4 Monte Carlo simulator is a list of the pixels having a non-zero energy deposit, with their
location and deposited energy. Only double events which release an amount of energy simultaneously
in both the LEID and the HEID are relevant to study polarization and are recorded. However, among
them we must select the good events, which are those for which it is possible to reconstruct
the direction of scattering, and therefore are useful to measure the polarization of the incident
photon beam. Examples of double but not good events are those in which one of the detectors collects
the fluorescence radiation emitted after the photoabsorption in the other detector or events which
undergo multiple scatterings in different pixels, especially in the LEID. Another process which
also produces double events is the photoabsorption in the few microns at the bottom surface of the
LEID: in this case the photoelectron may escape from the LEID and it could be absorbed in the
HEID. These events, albeit rare, must be removed because the direction of emission of the
photoelectron is sensitive to polarization but the effect is opposite to that of the Compton
scattering, i.e. the probability of emission is maximum along the polarization. In our simulations
we used a 30~$\mu$m thick beryllium plate between the LEID and the HEID to stop the large part of
the photoelectrons but not X-rays (the transparency of the plate at 20 keV is 99.9\% on-axis).
However the configuration of such a shield could be conveniently adapted to specific needs in actual
instruments, for example it could be a thin thermal shield if the LEID and the HEID must work at
different temperatures.

We defined some criteria to filter those events in which the photon underwent only two interactions,
the first as a scattering in the LEID and the second in the HEID, restricting ourselves only to
procedures which could be applied also to data collected by real instruments. The first filter is a
\emph{spatial filter}, and it is based on the fact that good events must undergo the first LEID
scatter in the same pixels in which the source is imaged. An accurate model of the instrument should
consider that the image of the source is distributed on the focal plane according to the point
spread function of the telescope. Instead, in our toy model we just choose a circular beam
distributed uniformly in the four central pixels of the detector. Therefore we select only those
events which release at least an energy deposit in one of these pixels and/or in their adjacent
ones. Secondly, we imposed a criterion on the number of hit pixels (\emph{hit pixels filter}). We
select only those events for which there is only one energy deposit in the LEID, which must be in a
location compatible with the spatial filter, and only one hit pixel in the HEID. The second
condition must be applied with some attention because we experienced that the charge produced by a
photoabsorption event may be spread over a cluster of several non-contiguous pixels, mainly because
fluorescence photons may be absorbed at some distance from the photoabsorption point. We treat such
a cluster as a unique pixel with a charge equal to the total energy deposit and a position weighted
with the energy collected in each pixel of the cluster. Operatively, we consider that all pixels
less distant than 10 pixels belong to the same cluster. Although such a large \emph{summation
radius} may imply some drawbacks on a real device, especially to remove the particle background,
large clusters must be correctly included in the analysis to avoid systematics on the modulation
curve due to the square pattern of the pixels (see Figure~\ref{fig:I_MC_UnP}). A third filter is
applied on the energy collected on the HEID, which must be higher than 5~keV (\emph{HEID energy
filter}). Such a filter has not a large impact on the efficiency ($\sim$3\% at 30~keV) or on the
modulation factor (variation below the statistical error). Notwithstanding, we imposed a energy
threshold for the HEID events in order to remove those double events in which the photon is
absorbed in the LEID and the fluorescence of Silicon at $\sim$1.7~keV in the HEID. The 5~keV value
is reasonable for current CdTe detectors. The last filter we defined, called \emph{$\theta$-filter},
is a filter on the scattering angle, measured by the position of the pixels hit in the two
detectors. We used the same limits derived in Section~\ref{sec:AnalyticalQF}, that is we restricted
the scattering angle between 30 and 52$^\circ$.

\begin{figure*}[htbp]
\begin{center}
\includegraphics[angle=90,width=\textwidth]{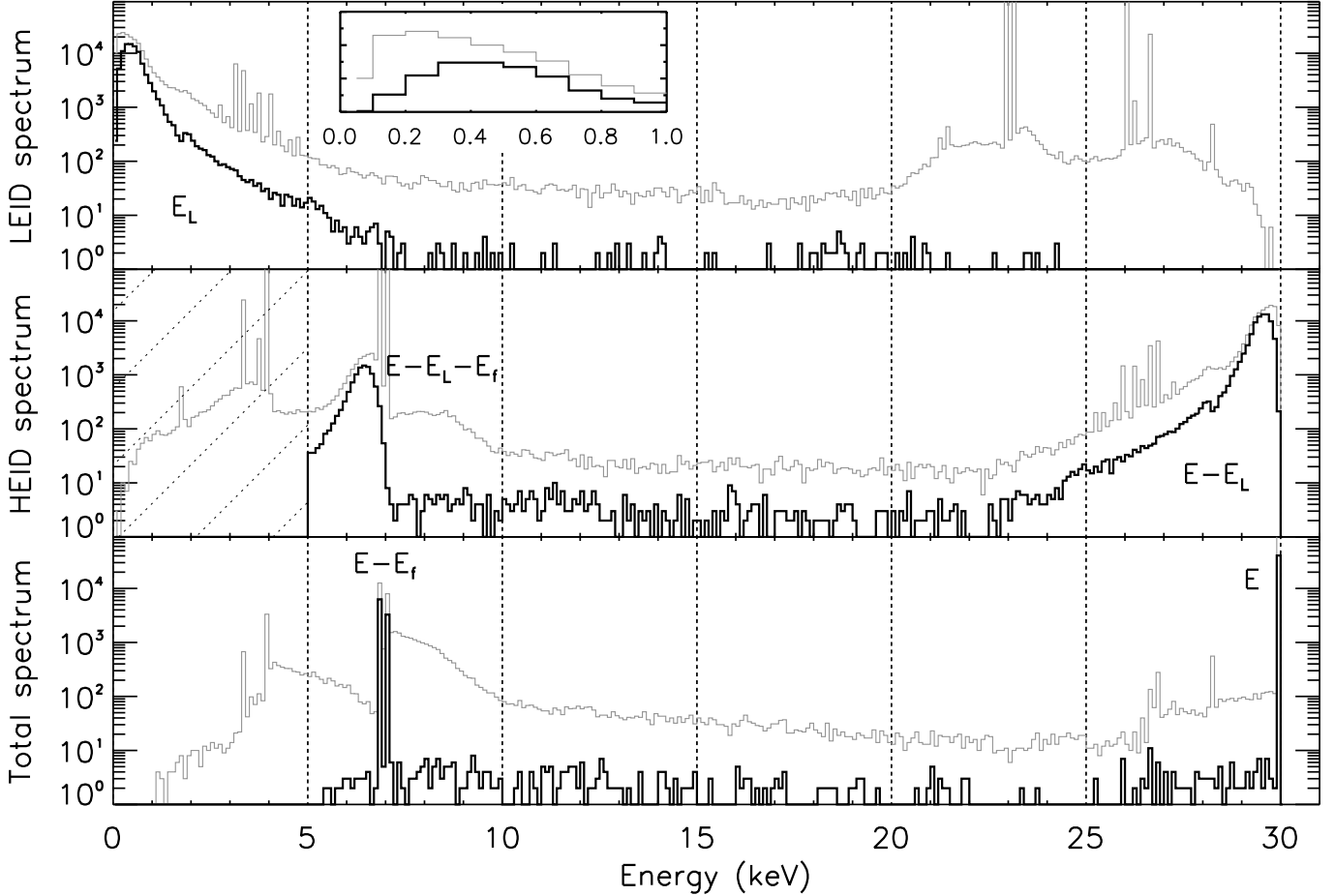}
\end{center}
\caption{Spectrum measured by the LEID (top) and by the HEID (middle) for 30~keV incident photons.
The total spectrum, obtained by summing the energy collected by the two instruments event by event,
is in the bottom panel. The inset in the top panel shows a zoom at low energy of the LEID spectrum.
In black and gray are reported the spectrum with and without the filters discussed in the text
to distinguish real scattering events, including the $\theta$-filter. The diagonally dashed region
in the middle panel shows the energy threshold imposed for the HEID with the energy filter.}
\label{fig:I_fffSE_030keV}
\end{figure*}

\begin{figure*}[htbp]
\begin{center}
\includegraphics[angle=90,width=\textwidth]{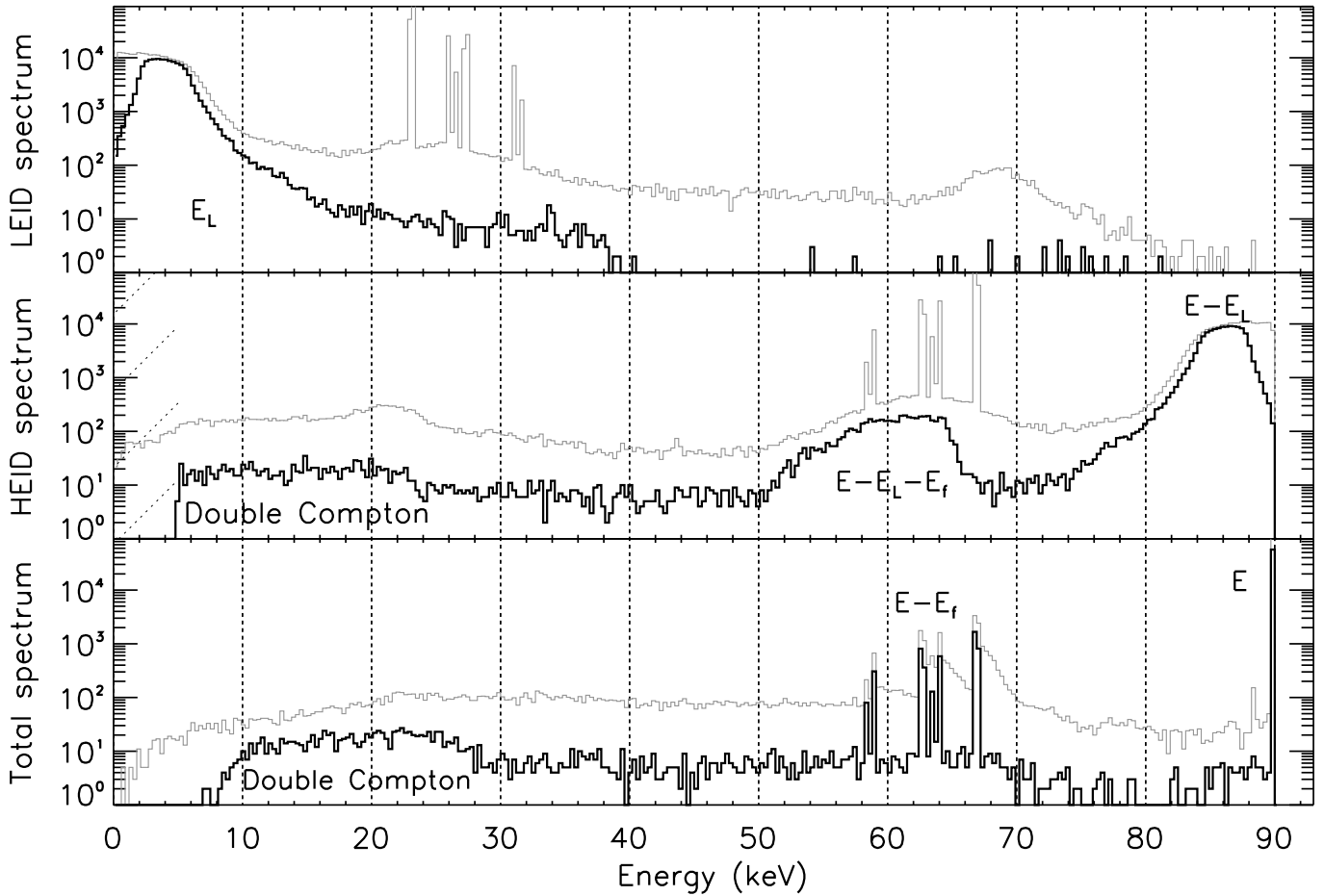}
\end{center}
\caption{The same as Figure~\ref{fig:I_fffSE_030keV} but for 90~keV photons.}
\label{fig:I_fffSE_090keV}
\end{figure*}

The spectrum obtained for the LEID, the HEID and by summing event by event the energy deposits in
the two detection planes is shown in Figure~\ref{fig:I_fffSE_030keV} for 30~keV incident photons.
The spectrum including all events is reported as a gray histogram, while that obtained after
applying the filters discussed above is in black. A lot of prominent lines are visible and
correspond to the escape peaks or to the absorption of fluorescence lines from Silicon (K-lines at
$\sim$1.7~keV), Cadmium (K-lines at $\sim$23\--26~keV, L lines at $\sim$3~keV) or Tellurium (K-lines
at $\sim$27\--31~keV, only visible for photons above $\sim$32~keV, and L lines at $\sim$4~keV). Good
events are basically characterized by a small energy deposit $E_{L}$ in the LEID (a few keV) and by
an energy deposit in the HEID equal to $E-E_{L}$, where $E$ is the energy of the incident photons.
These events fill in the total spectrum the energy bin around 30~keV and represent by far the main
components after applying the filters (black histogram in Figure~\ref{fig:I_fffSE_030keV}). Another
class of good events are those which deposit in the HEID an energy equal to $E-E_{L}$ but the
fluorescence photon, which is emitted with high probability ($>$80\% for K-shell
photoabsorption and high-Z material like Cadmium and Tellurium), escapes from the detector. In this
case the energy detected in the HEID is $E-E_{L}-E_{f}$, where $E_f$ is the energy of the
fluorescence photon, and therefore these events make up an escape peak in the total spectrum at
energy $E-E_f$. In Figure~\ref{fig:I_fffSE_030keV} two of such escape peaks are visible at about
7~keV, which correspond to the escape of Cd~K$_{\alpha 1}$ and Cd~K$_{\alpha 2}$ fluorescence
photons with energy 23.2~keV and 23.0~keV, respectively. It is worth mentioning that the application
of the $\theta$-filter has the positive effect to make our results less dependent on the energy
threshold of the LEID. As a matter of fact, the removal of events scattered at less of 30$^\circ$
cut out the large part of the events which deposit in the LEID only a few hundreds of eV (see the
inset in top panel of Figure~\ref{fig:I_fffSE_030keV}). For the sake of completeness, we also show
in Figure~\ref{fig:I_fffSE_090keV} the spectrum for 90~keV incident photons. The main features
remain the same as those discussed in the case of 30~keV radiation, although in this case it is
visible a low energy bump in the HEID and in the total spectrum which corresponds to photons which,
after the scattering in the LEID, scatter a second time in the HEID. Such events are still good to
measure polarization, although their statistical weight is negligible with respect to the events
which are absorbed after the first scattering.

Events which pass all the filters described above are used to build a modulation curve which is
fitted with the function
\begin{equation}
\mathcal{M}(\varphi) = A + B\cdot\cos(\varphi-\varphi_0)\, .
\end{equation}
The modulation factor $\mu$ is derived as usual by (cfr. Equation~\ref{eq:MuDef})
\begin{equation}
\mu =
\frac{\mathcal{M}_\mathrm{max}-\mathcal{M}_\mathrm{min}}{\mathcal{M}_\mathrm{max}+\mathcal{M}
_\mathrm{min}} =\frac{B}{ B+2A } .
\end{equation}

\subsection{Results} \label{sec:Results}

We performed simulations at 20, 30, 40, 50, 60, 70, 80, 90~keV and 100~keV, generating $50\cdot10^6$
photons for each energy. Taking as an example the simulation at 30~keV, we detected $1.2\cdot10^6$
double events, that are for the large part events in which the photon is absorbed in the HEID and
some fluorescence in the LEID. Among them, $162\cdot10^3$ events were selected by the spatial
filter, $156\cdot10^3$ passed both the spatial and the hit pixels filter, $151\cdot10^3$ passed also
the HEID energy filter and eventually $93\cdot10^3$ passed also the $\theta$-filter. We repeated the
simulations for three different values of the angle of polarization, that are 0$^\circ$, 25$^\circ$
and 90$^\circ$, to check the consistency of our results. In this respect, the first and the last
values represent the two extreme conditions, while 25$^\circ$ is just an intermediate value not in
resonance with any expected on a square pattern.

\begin{figure}[htbp]
\begin{center}
{\includegraphics[angle=90,width=7.5cm]{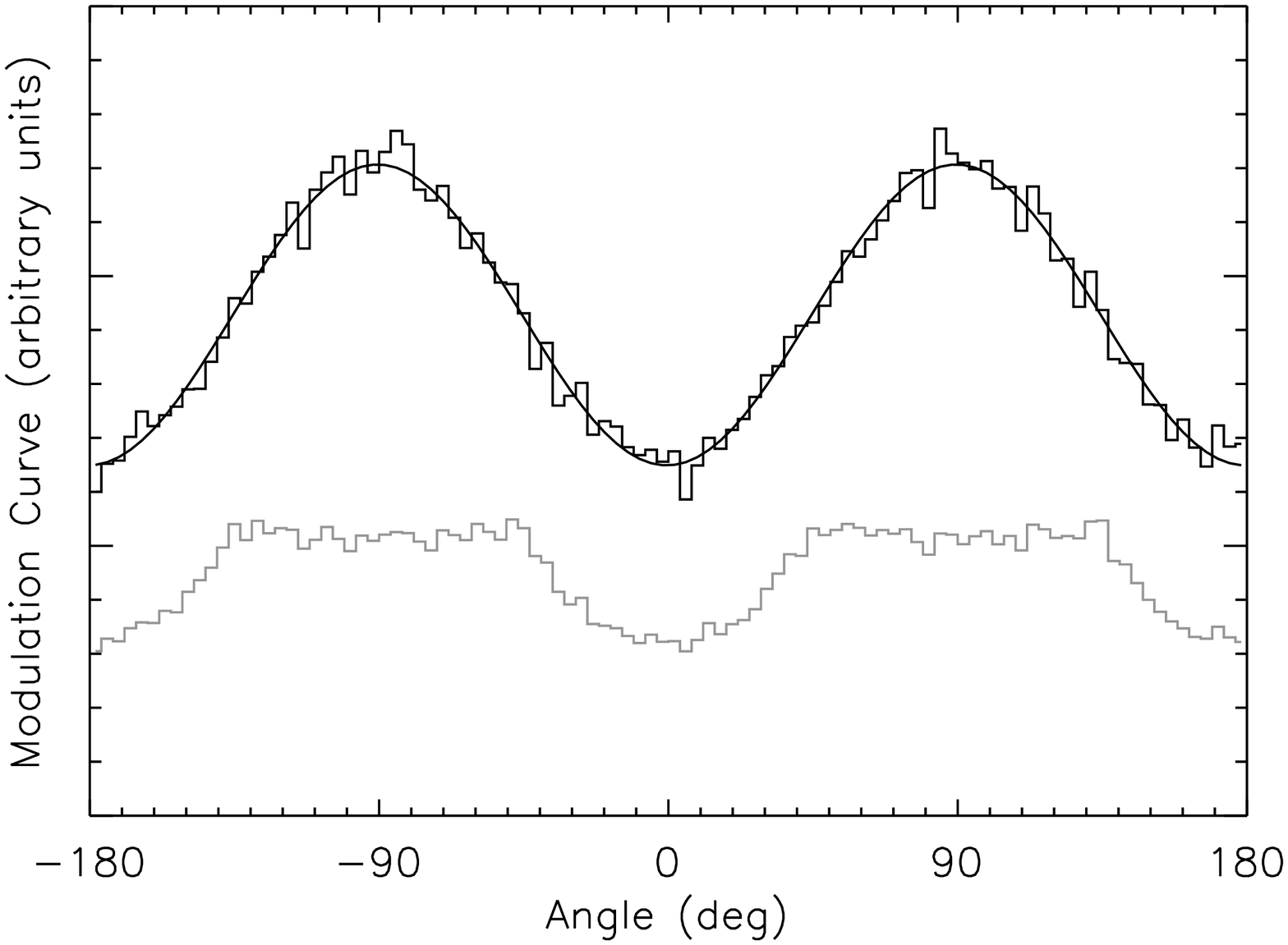}}
\hspace{5mm}
{\includegraphics[angle=90,width=7.5cm]{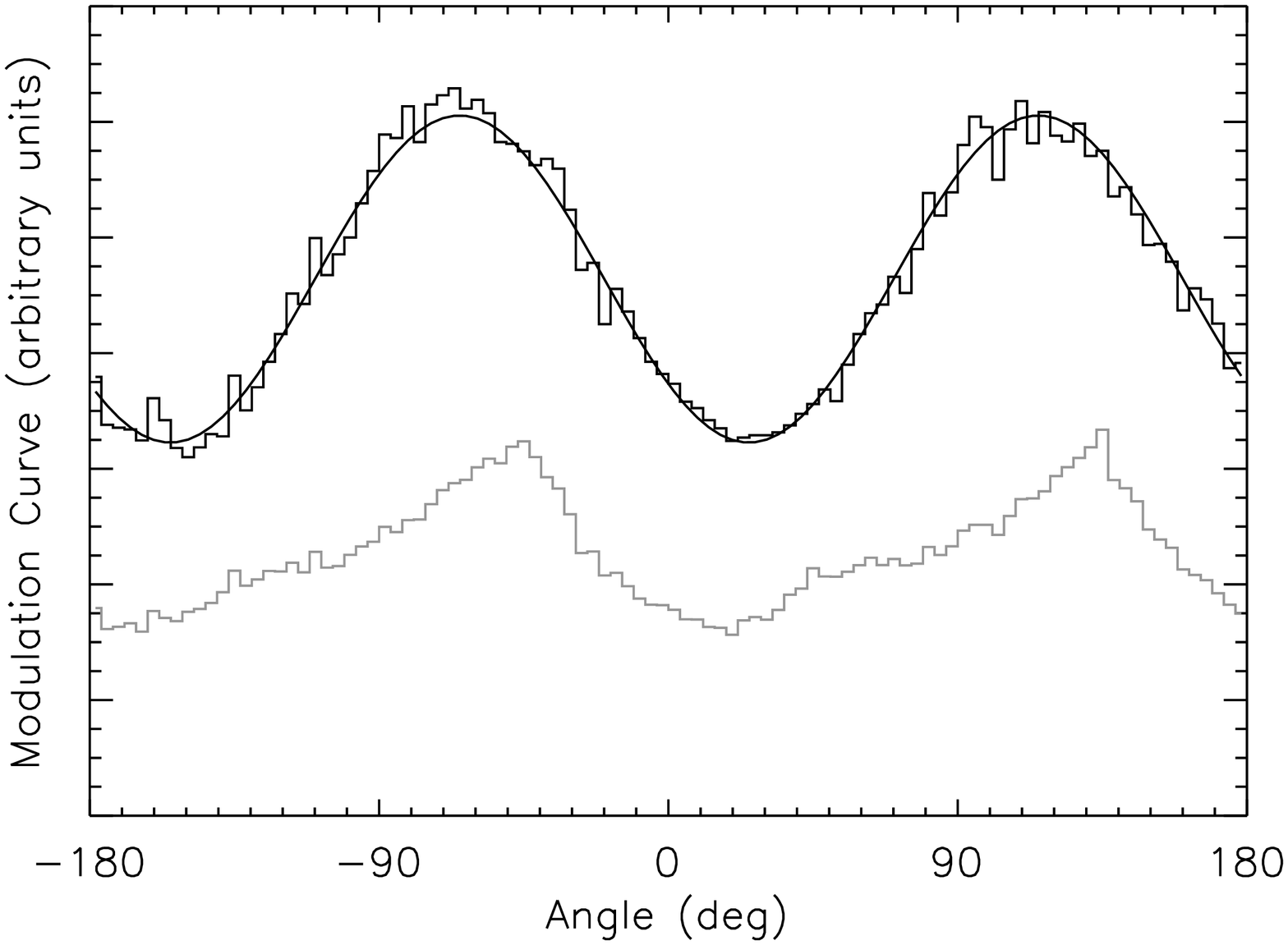}}
\end{center}
\caption{Modulation curve obtained from Monte Carlo simulations for an angle of
polarization 0$^\circ$ (upper panel) or 25$^\circ$ (lower panel). The black and gray histograms refer to the
modulation curve obtained if the $\theta$-filter is applied or not, respectively. The black and
gray histograms are not to scale.}
\label{fig:I_MC}
\end{figure}

Examples of the typical modulation curves obtained from Monte Carlo are shown in
Figure~\ref{fig:I_MC} in the case of 30~keV photons and a polarization angle of 0$^\circ$ or
25$^\circ$. The fit with the square-cosine function is very good and the reduced $\chi^2$ is 1.067
and 1.042 respectively for 97 degrees of freedom. We report in the same Figure as the gray histogram
also the modulation as it would appear without applying the $\theta$-filter. In this case a strong
systematic effects appears in the modulation curve which is simply due to the fact that the HEID
detection plane is square. As a matter of fact, in each bin of the modulation curve there is the
number of events scattered in a certain azimuthal angular bin regardless the scattering angle.
Without the $\theta$-filter, the scattering angle is only constrained by the geometry of the two
detection planes because the photon must cross the HEID to be detected. However, the detection
planes are square, and this implies that the interval of accepted scattering angles is larger in the
diagonal directions thus causing a bump in the number of collected events.

\begin{figure}[htbp]
\begin{center}
{\includegraphics[angle=90,width=7.5cm]{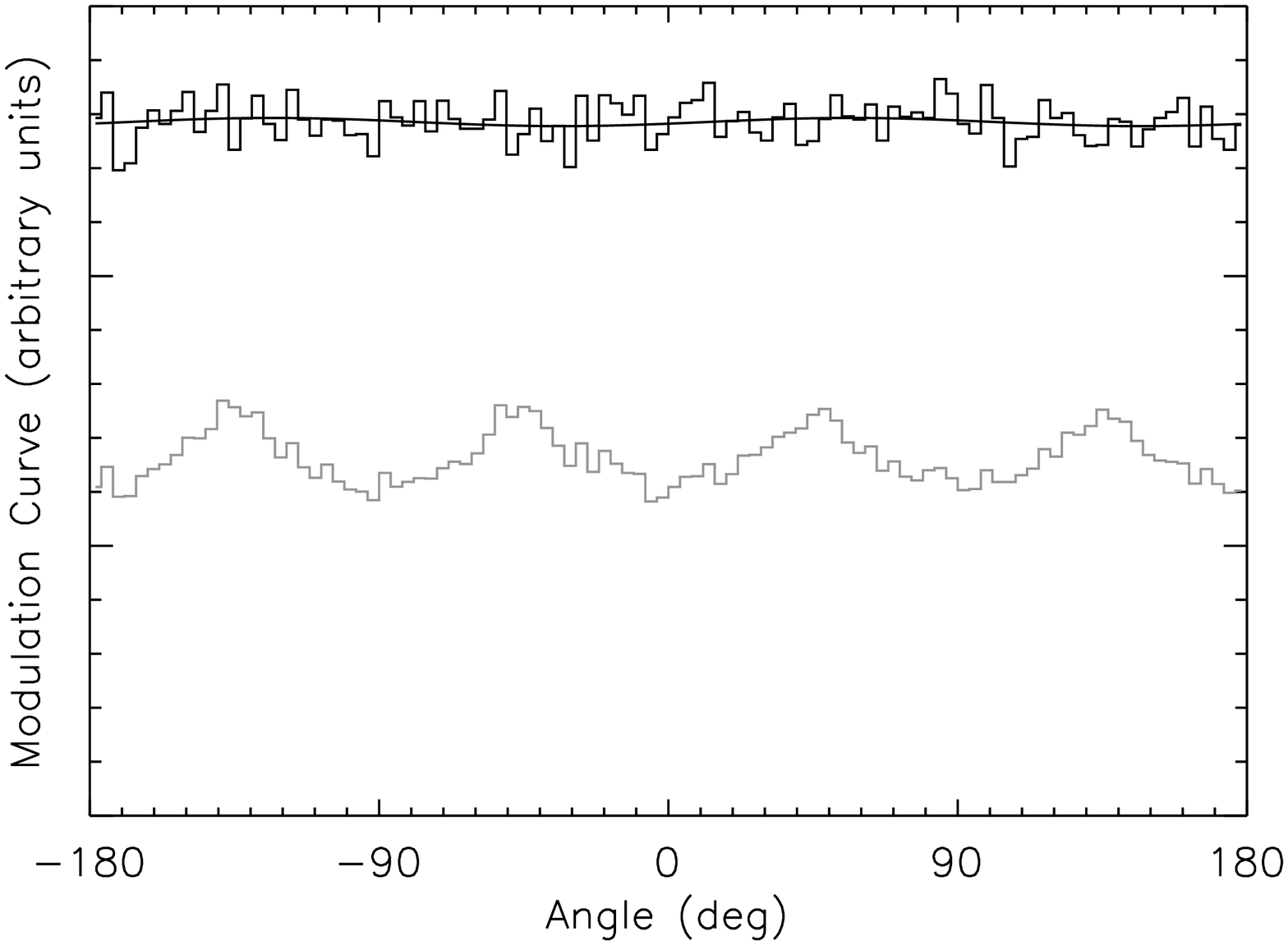}}
\hspace{5mm}
{\includegraphics[angle=90,width=7.5cm]{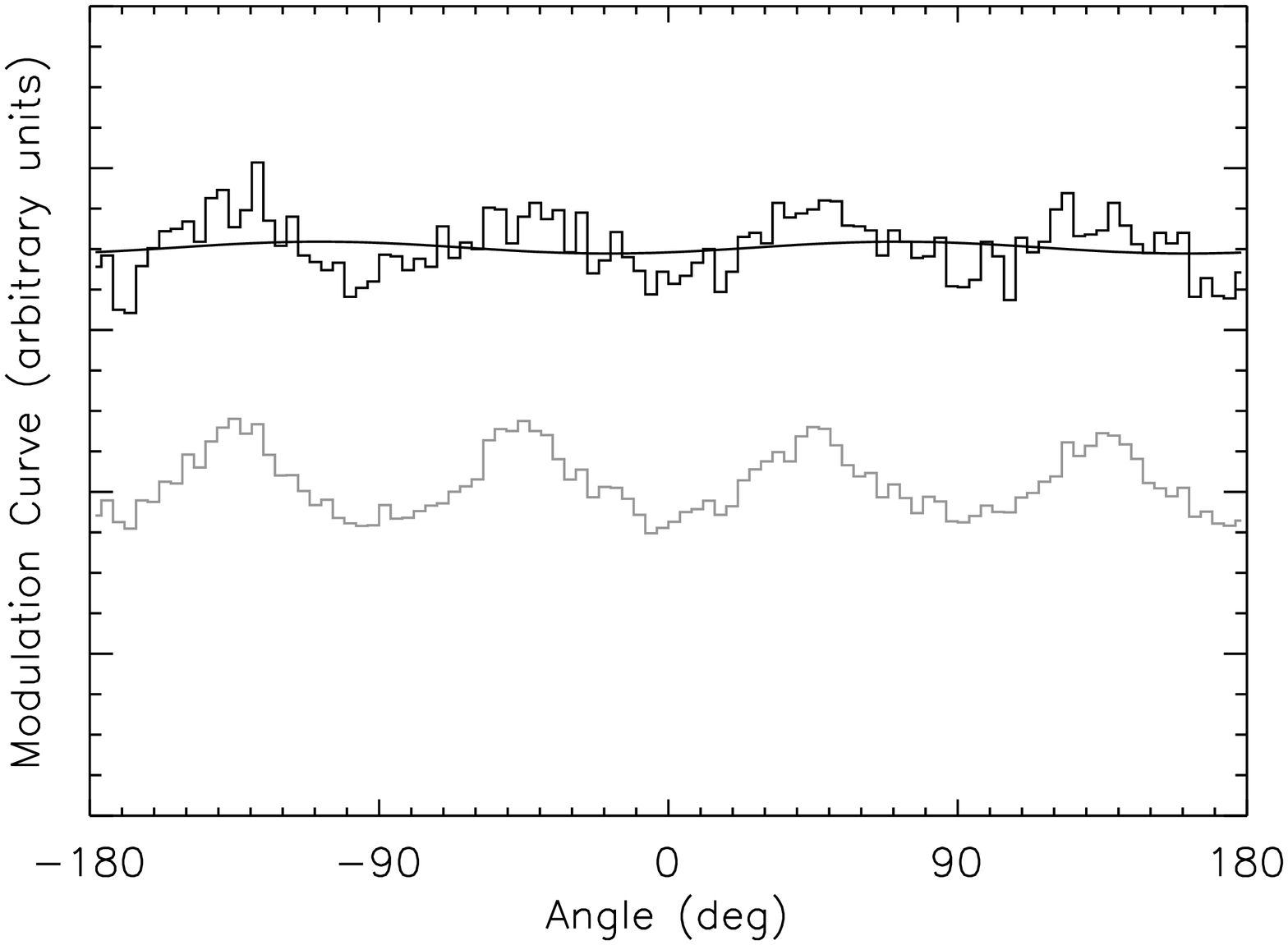}}
\end{center}
\caption{Modulation curve obtained for unpolarized radiation at 90~keV. Upper panel: Modulation curve
obtained from Monte Carlo simulations if the $\theta$-filter is applied or not (black and gray
histogram respectively). The two histograms are not to scale. Lower panel: The same as the upper panel,
but in the case the summation radius is 1.6 instead of 10, that is only contiguous pixels are
considered as an unique cluster.}
\label{fig:I_MC_UnP}
\end{figure}

Although we neglected all of the main causes which may induce a systematic modulation for a real
instrument, it is interesting to look at the behavior of the modulation curve for completely
unpolarized photons. This is reported in Figure~\ref{fig:I_MC_UnP} for 90~keV photons, in
which as above the black and gray histograms are those obtained with or without applying
the $\theta$-filter, respectively. As expected, the detected modulation is consistent with zero with
a 33\% confidence level. However, it is also interesting to see what would be the result of our
analysis in case of slightly different filter parameters. When we described the hit pixels filter
in the Section~\ref{sec:DataAnlysis}, we mentioned that a large summation radius, that was 10
pixels, is required to avoid systematic effects on the modulation curve. In the lower panel of
Figure~\ref{fig:I_MC_UnP} we show the modulation curve obtained by exactly the same data as in
the upper panel, but with the only difference that the summation radius is 1.6 pixels
instead of 10. In this case only pixels which are contiguous on the side or on the diagonal are
considered as a unique cluster and therefore only events involving a single cluster of contiguous
pixels pass the hit pixels filter. The result is that a strong systematics due to the square pattern
of the pixels emerges with a peak to peak variation of $\sim$10\%. Although formally the
square-cosine modulation is still consistent with unpolarized radiation at a 15\% confidence level,
this is only due to the high symmetry of the four peaks. In actual devices also small
nonuniformities would cause a much higher modulation and in the most extreme situation the spurious
polarization, obtained by dividing the peak to peak variation by the modulation factor, would be
$\sim$10\%/$\mu\approx30$\%. As a matter of fact, spurious modulated signals can arise from minor
effects even in apparent symmetrical geometries.

\begin{figure*}[htbp]
\begin{center}
\includegraphics[angle=90,width=11cm]{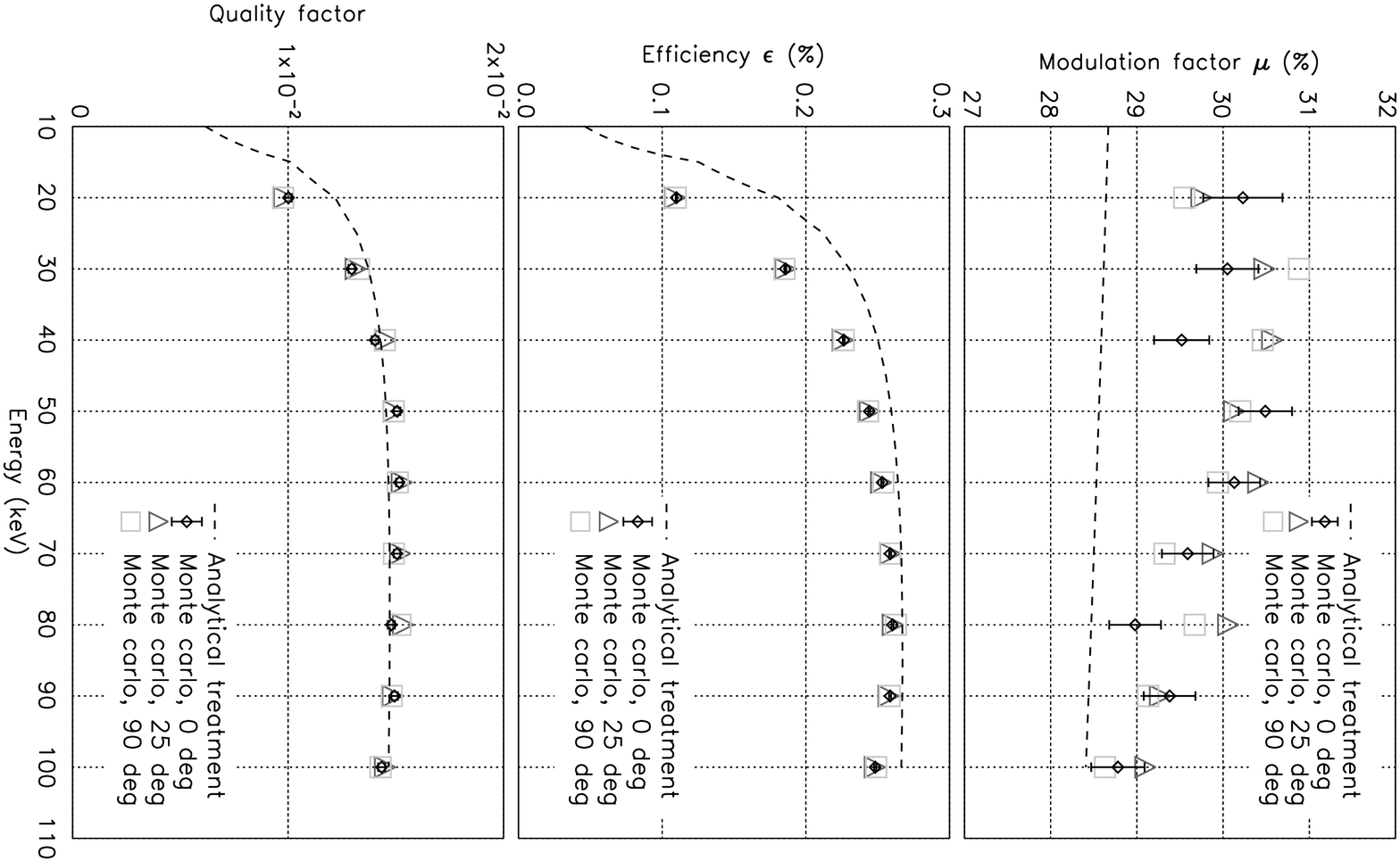}
\end{center}
\caption{Comparison of the modulation factor (top), efficiency (middle) and quality factor (bottom)
as derived by the analytical treatment (dashed line) and by the Monte Carlo simulations (points).
The different symbols refer to different values of the incident photons angle of polarization,
0$^\circ$, 25$^\circ$ or 90$^\circ$. Error bars are shown only for the first value of the
polarization angle for graphical reasons, but they are representative for the other values too. The
small disagreement between the analytic and Monte Carlo results, which decreases with energy, is due
to the fact that in Monte Carlo simulations we treated more accurately the physics of the Compton
scattering.}
\label{fig:I_MuEffQF} 
\end{figure*}

\begin{figure*}[htbp]
\begin{center}
{\includegraphics[angle=90,width=7cm]{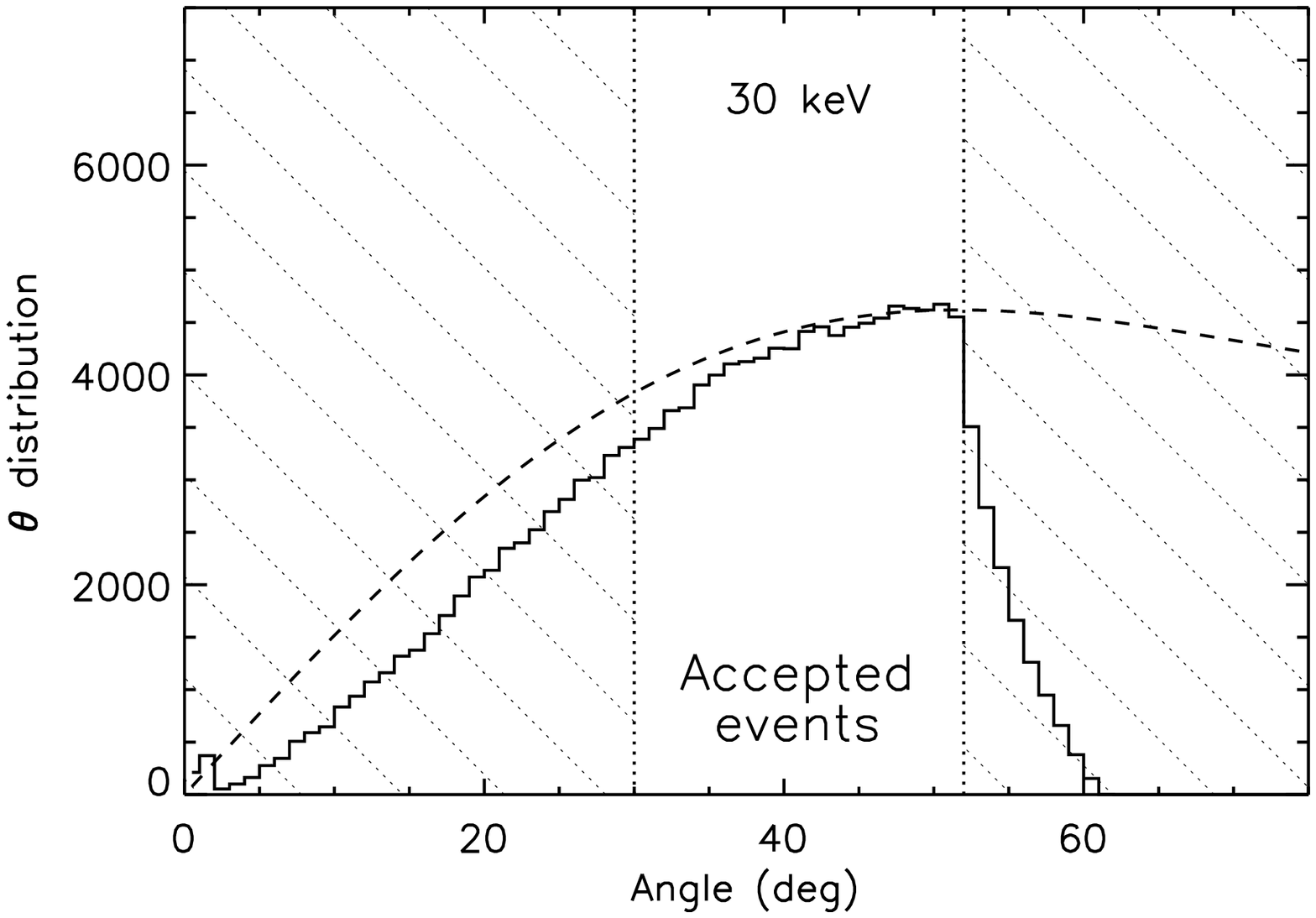}}
\hspace{5mm}
{\includegraphics[angle=90,width=7cm]{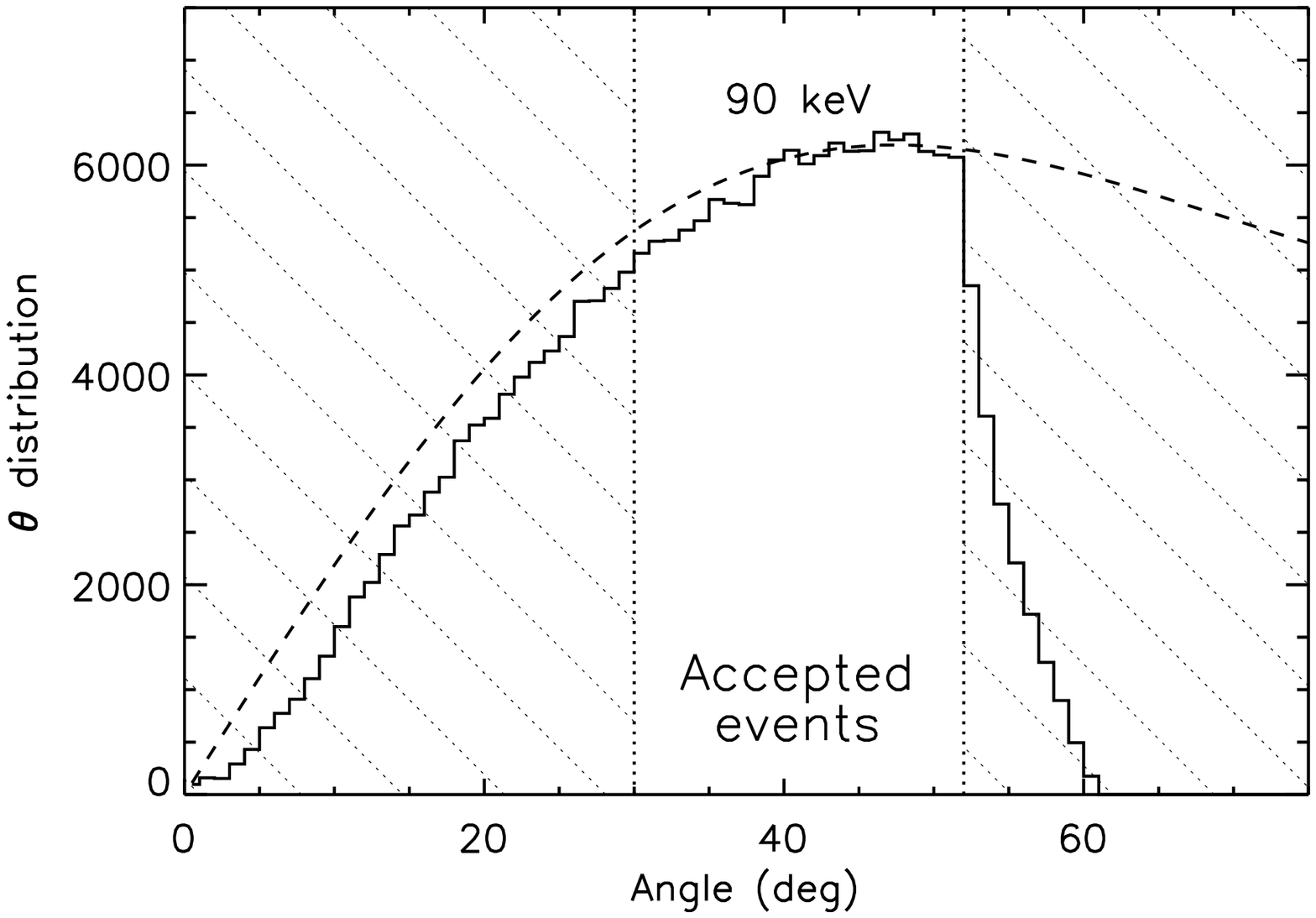}}
\end{center}
\caption{Comparison of the scattering angle distribution for the events which passed all the
filters in Monte Carlo simulations at two energies, 30~keV (left) and 90~keV (right). The
distributions are compared with that expected on the basis of the Klein-Nishina formula used in the
analytical treatment (dashed line). The difference between the two distributions originate the small
disagreement visible in Figure~\ref{fig:I_MuEffQF}. The diagonally dashed areas are the regions
which are cut out by the $\theta$-filter.}
\label{fig:I_TD}
\end{figure*}

In Figure~\ref{fig:I_MuEffQF}, the modulation factor, the efficiency and the quality factor obtained
from Monte Carlo are compared to the values derived by the analytical treatment discussed in
Section~\ref{sec:Analytical}. The efficiency is calculated as the ratio between the number of
events which passed all filters and that of incident photons. The agreement is in general rather
good, although there are some deviations with respect to the expected dependence. In particular, the
modulation factor is larger and the efficiency is smaller than expected, with a difference which
decreases with energy. This disagreement is easily explainable in view of the fact that in Monte
Carlo simulations we treated Compton scattering more accurately by including the scattering
function. As a matter of fact, this has the effect to suppress the forward scattering, especially at
low scattering angles and at low energy. Only forward scatterings are collected in the stacked
imager layout and therefore an efficiency reduction is a straightforward consequence of introducing
the scattering function. This can as well explain the increase of the modulation factor because the
suppression is more effective at low scattering angles, where the intrinsic modulation with
polarization is lower. This interpretation is confirmed by studying the scattering angle
distribution of the events which passed the filters. At low energy (see
Figure~\ref{fig:I_TD}, left panel) there is a significant difference between the distribution obtained
from Monte Carlo and that expected from the Klein-Nishina formula, especially at low scattering
angles. This explains why the estimates of the modulation factor and of the efficiency obtained
analytically and with the Monte Carlo are different. Instead the results are more in agreement at
higher energy, where the two distributions differ less (see Figure~\ref{fig:I_TD}, right panel).

\begin{figure}[htbp]
\begin{center}
\includegraphics[angle=90,width=8cm]{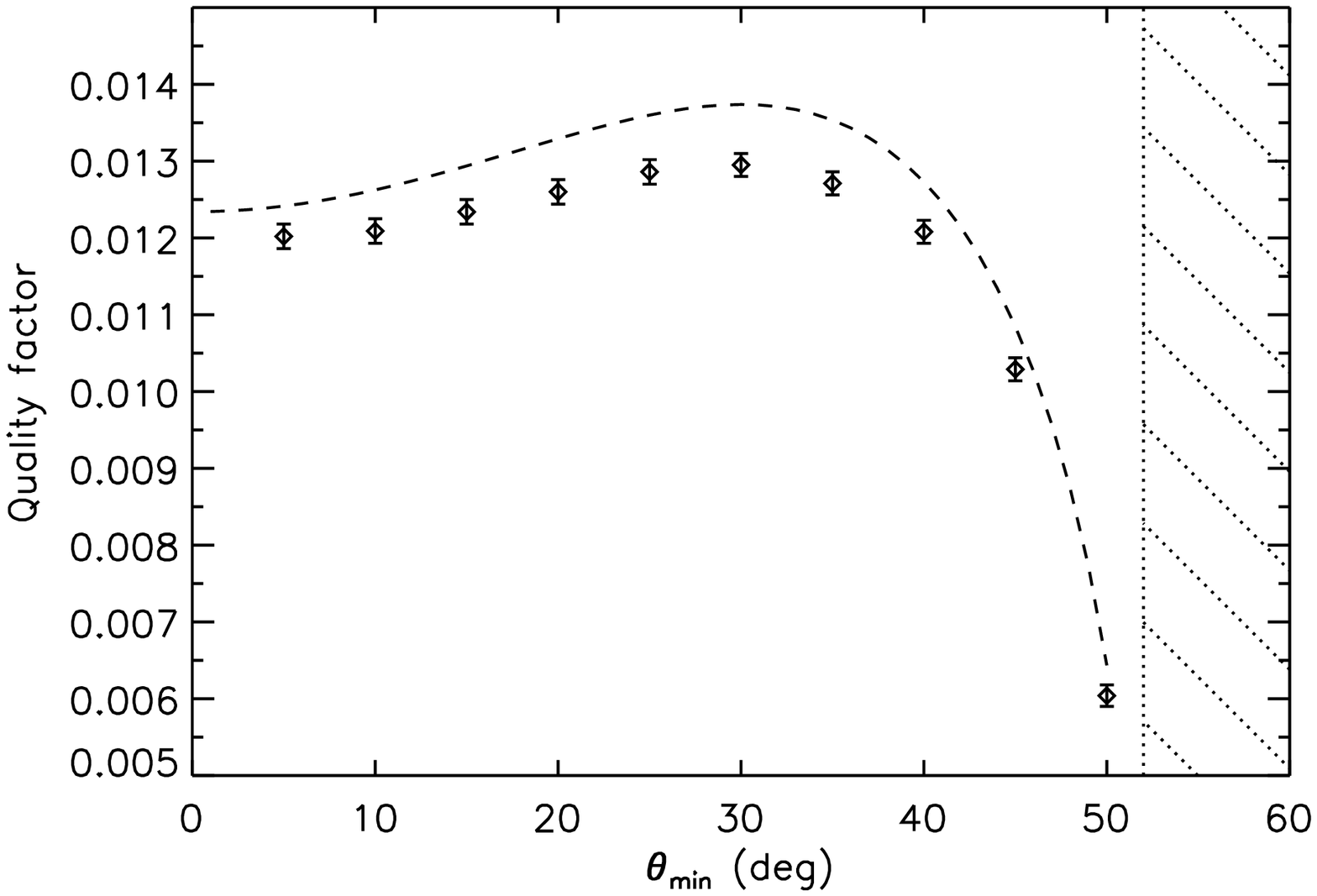}
\end{center}
\caption{Quality factor as a function of $\theta_{\mathrm{min}}$ for
$\theta_{\mathrm{max}}=52^\circ$ and 30~keV photons. The dependence derived by the Monte Carlo
(points) is in very good agreement with that expected from our analytical treatment and confirm a
maximum sensitivity for $\theta_{\mathrm{min}}=30^\circ$. The small offset can be ascribed
as above to the fact that in Monte Carlo simulations we included the scattering function.}
\label{fig:I_QFvsThetaMin} 
\end{figure}

Monte Carlo simulations allow us to confirm another result obtained in the analytical treatment
presented in Section~\ref{sec:Analytical}, that is the maximum of the sensitivity for
$\theta_{\mathrm{min}}\approx30^\circ$ (see Equation~\ref{eq:QFmax}). In
Figure~\ref{fig:I_QFvsThetaMin} we show the quality factor as a function of $\theta_{\mathrm{min}}$
for $\theta_{\mathrm{max}}=52^\circ$ and 30~keV photons. The dependence obtained from the
simulations is closely related with the expected one, with a small offset which can be again
ascribed to the inclusion of the scattering function in the simulations.

\section{Conclusions}

The primary aim of this work was to assess the intrinsic sensitivity to polarization of the stacked
imager design more than to discuss the feasibility and the performance of a particular
implementation. In this respect, the results of our analytical treatment and of Monte Carlo Geant4
simulations are in full agreement. The first noteworthy result is the discovery of a peak in the
quality factor, a quantity assumed to be representative of the polarization sensitivity, for a
particular range of accepted scattering angles.  It is worth mentioning that this non-obvious
result firstly came out by our analytical treatment. This suggests that an optimization of Compton
scattering polarimeters can (and should) be done also with elementary considerations based on the
Klein-Nishina cross section. Unfortunately, our main result is that the stacked imager design has a
very low sensitivity to polarization despite our optimistic assumptions. We are far from an
optimized layout and some improvement is still possible with a more refined data analysis. For
example, in the stacked imager design the appropriate modulation factor could be associated to each
event depending on the scattering angle instead of using an average value.  Also, it would be
possible to add more detectors encircling the imagers in order to detect photons scattered on a
large fraction of the solid angle and not only in the forward direction. However, all of these
tricks may improve the sensitivity of a factor of a few at maximum while the low sensitivity of the
stacked imager design is largely limited by the very low probability for scattering in the first
detection plane. At this regard, we do not think that a major improvement is possible since our
estimates are based on the fundamental physics of the Compton scattering. An increase in the
scattering efficiency, even though not outstanding, would be possible only by increasing the Silicon
thickness. However this would unavoidably worsen the performance of the detector as an imager and
this may be unacceptable, since the main use of such a stacked imaging detector would be in any case
to image photons on a large energy range. In addition, the dependence of the quality factor on the
efficiency is rather weak and, roughly speaking, doubling the efficiency would result in just a 40\%
increase in quality factor.

Obviously, a certain polarization sensitivity, albeit low, is present in any case and the common
argument is that it can be exploited ``for free''. However we think that this argument should be
discussed with care because the choice to use stacked imagers to detect scattered photons poses
some requirements which are usually not set if the instrument is used just as an imager. For
example, fast coincidence between the two detection planes is an ad-hoc requirement. Specific
calibration campaigns with polarized and unpolarized radiation would be required in order to
obtain reliable results and to single out e.g. nonuniformities in the pixel response. Dedicated
efforts would also be needed for data analysis. In the most optimistic assumption of negligible
background, good events are still overwhelmed by other double events by an order of magnitude, and
the application of some of the filters that we used, e.g. the hit pixel filter, may be difficult for
real instruments. Another issue which we intentionally disregarded but requires significant efforts
when dealing with real instruments is the control of the systematics which may be severe for the
stacked imager layout.

The low sensitivity we estimated is also not encouraging when it is compared with the performance
of dedicated instruments. For example, the focal plane Compton polarimeter discussed by
\citet{Soffitta2010} (but see also Fabiani et al. 2012 in preparation) has a quality factor which is
higher than 0.40 at 30~keV and almost constant with energy, i.e. at least a factor $\sim$30
higher than that of a stacked imager polarimeter. Roughly speaking, this means that the same MDP
can be reached with an observation $\sim30^2=900$ times shorter. This obviously discourages the
efforts to develop and calibrate the polarization capabilities of forthcoming stacked imagers and
rather suggests to use dedicated instruments. A possible policy which could be adopted in case of
missions with a large telescope is to mount the stacked imager to perform imaging and a dedicated
polarimeter on a movable platform and to place alternatively the two instruments in the focus of the
telescope. For example, this was the strategy adopted for XEUS/IXO. If the science objectives of the
mission require simultaneous spectral-imaging and polarimetric measurements, a solution could be
to use two different telescopes for the two instruments. After all, even used with a telescope with
a collecting area 10 times smaller than that of the stacked imager, the dedicated polarimeter would
still require an observation time $\sim$90 times shorter to reach the same MDP.

As a final note, it is worth mentioning that our results are \emph{not} in agreement with those
reported by other authors. In particular, \citet{Gouiffes2008} reported for the stacked imagers
on-board Simbol-X a 35\%~MDP for 100~mCrab sources in 10~ks between 20 and 80~keV, which corresponds
to 11\% MDP for 100~mCrab sources in 100~ks. Using the values derived by the Monte Carlo simulations
(see Fig.~\ref{fig:I_MuEffQF}) and without adding any background, we derived a MDP $\sim$28\% in the
latter case, that is a factor $\sim(28/11)^2\approx6$ worse in observation time. For the COSPIX
mission, \citet{Ferrando2010} reported a 0.7\%~MDP for 100~mCrab sources in 100~ks between 20~keV
and 40~keV, while our estimate is 18\%. Incidentally, a dedicated polarimeter such as that presented
by \citet{Soffitta2010} in the focus of the COSPIX telescope would reach a MDP 0.3\% for 100~mCrab
sources in 100~ks. It is difficult to establish the origin of the disagreement between our results
and those presented by previous authors because the value of the modulation factor, of the
efficiency and of the background assumed to derive the MDP were not declared. Nevertheless, these
disagreements can not be imputed only to the different geometry of the assumed detector,  since our
analysis suggests a rather weak dependency on parameters such as the distance between the detection
plane or the size of the detectors.

\begin{acknowledgements}
FM and RC acknowledges the financial support on the INAF contract PRIN-INAF-2009. The authors
are grateful to the anonymous referee for his/her interesting comments to this work.
\end{acknowledgements}

\bibliography{References}
\bibliographystyle{apj}

\end{document}